\documentclass[conference,a4paper]{IEEEtran}
\usepackage[utf8]{inputenc}
\usepackage[normalem]{ulem}
\usepackage{amsmath}
\usepackage{amsfonts}
\usepackage{amssymb}
\usepackage{algpseudocode}
\usepackage{algorithm}
\usepackage[pdftex]{graphicx}
\usepackage{epstopdf,epsfig}
\usepackage{enumerate}
\usepackage{pgfplots}
\usepackage{bbold}
\usepackage{subfigure}
\usepackage{mathtools}
\usepackage{amsthm}
\usepackage{wrapfig}
\usepackage{caption}
\usepackage[english]{babel}
\usepackage{url}
\usepackage{color}

\ifCLASSOPTIONcompsoc
  \usepackage[nocompress]{cite}
\else
  \usepackage{cite}
\fi

\theoremstyle{remark}

\theoremstyle{theorem}

\newtheorem{theorem}{Theorem}[section]
\newtheorem{lemma}[theorem]{Lemma}
\newtheorem{corollary}[theorem]{Corollary}

\theoremstyle{definition}
\newtheorem{definition}{Definition}[section]

\newcommand{\karmoose}{\color{black}}

\newenvironment{Pf}{\textbf{Proof:}}{\hfill$\blacksquare$}

\newcommand{\Set}{\mathcal}
\title{Privacy in Index Coding: Improved Bounds and Coding Schemes}
\author{
\IEEEauthorblockN{Mohammed Karmoose$^\dagger$, Linqi Song$^\dagger$, Martina Cardone$^\ddagger$, Christina Fragouli$^\dagger$}
$^\dagger$University of California Los Angeles, Los Angeles, CA 90095 USA\\
$^\ddagger$University of Minnesota, Minneapolis, MN 55455 USA\\
Email: \{mkarmoose, songlinqi\}@ucla.edu, cardo089@umn.edu, christina.fragouli@ucla.edu
}
\IEEEoverridecommandlockouts

\begin{document}

\maketitle
\begin{abstract}
It was recently observed in~\cite{karmoose2017preserving}, that in index coding, learning the coding matrix used by the server can pose privacy concerns: curious clients can extract information about the requests and side information of other clients. One approach to mitigate such concerns is the use of $k$-limited-access schemes~\cite{karmoose2017preserving}, that restrict each client to learn only part of the index coding matrix, and in particular, at most $k$ rows. 
These schemes transform a linear index coding matrix of rank $T$ to an alternate one, such that each client needs to learn at most $k$ of the coding matrix rows to decode its requested message.  This paper analyzes  $k$-limited-access schemes. First, a worst-case scenario, where the total number of clients $n$ is $2^T-1$ is studied. For this case,  a novel construction of the coding matrix is provided and shown to be order-optimal in the number of transmissions. 
Then, the case of a general $n$ is considered and two different schemes are designed and analytically and numerically assessed in their performance. It is shown that these schemes perform better than the one designed for the case $n=2^T-1$.
\end{abstract}

\section{Introduction} 
\label{sec::introduction}
It is well established that coding is necessary to optimally use wireless broadcasting for information transfer. The index coding framework, in particular, exemplifies the benefits of coding when using broadcast channels. In fact, by leveraging their side information, the requests of multiple clients can be simultaneously satisfied by a set of coded broadcast transmissions, the number of which could potentially be much smaller than uncoded information transfer~\cite{bar2011index}.
 
However, as we observed in~\cite{karmoose2017private,karmoose2017preserving}, coding also poses privacy  concerns: by learning the coding matrix, a curious client can infer information about the identities of the side information and request of other clients. In this paper, we build on the work in \cite{karmoose2017private,karmoose2017preserving} with the goal to offer improved constructions and bounds that enable to balance the trade-off between privacy and efficient broadcasting.

In an index coding setting, a server with $m$ messages is connected to $n$ clients via a lossless broadcast channel. Each client requests a specific message and may have a subset of the messages as side information. To satisfy all clients with the minimum number of transmissions $T$, the server can send coded broadcast transmissions; the clients then would use the coding matrix to decode their requests. In~\cite{karmoose2017preserving}, we  mitigated the aforementioned privacy risk by providing clients with access not to the entire coding matrix, but {\it only}  to the rows required for them to decode their own requests. 
In fact, given a coding matrix that uses $T$ transmissions to satisfy all clients, we can transform it into another coding matrix that uses $T_k\geq T$ transmissions to satisfy all clients, but where each client needs to learn only $k$ rows of the coding matrix.  In~\cite{karmoose2017preserving}, we showed that the attained amount of privacy is dictated by $k$.

This formulation admits a geometric interpretation. 
In~\cite{bar2011index}, it was shown that designing an index code is equivalent to the rank minimization of an $n\times m$ matrix $\mathbf{G}$, where the $i$-th row of $\mathbf{G}$ has certain properties which enable client $i$ to recover its request. Assume that the rank of $\mathbf{G}$ is $T$; then, we can use as a coding matrix $\mathbf{A}$ any basis of this $T$-dimensional space. By doing so, client $i$ can linearly combine some vectors of $\mathbf{A}$ to reconstruct the  $i$-th row of $\mathbf{G}$.
The geometric interpretation of our problem is therefore the following: 
Given $n$ distinct vectors in a $T$-dimensional space, represented as the rows $\mathbf{G}$, we wish to find an overcomplete basis $\mathbf{A}_k$ of dimension $T_k\geq T$, such that each of the $n$ vectors can be expressed as a linear combination of at most $k$ of the $\mathbf{A}_k$ vectors.

In~\cite{karmoose2017preserving}, we formalized the intuition that the achieved level of privacy can increase by decreasing the number $k$ of rows of the coding matrix that a client learns.  We also derived  upper and lower bounds on $T_k$, with the former being independent of  $n$.
 In this paper, our  main contributions are as follows:
\begin{enumerate}
\item We derive an improved upper bound that again applies for all values of $n$, and show that, in contrast to the one in~\cite{karmoose2017preserving}, it is  order-optimal.
Our upper bound is constructive, i.e., it provides a concrete construction of a coding matrix.
\item For general $n \leq 2^T-1$, the previous construction does not always offer benefits over uncoded transmissions. For such cases, we propose two novel algorithms and assess their analytical and numerical performance. In particular,
we show their superior performance over other schemes through numerical evaluations.
\end{enumerate}

The paper is organized as follows. Section~\ref{sec::setup} formulates the problem and  presents existing results. Section~\ref{sec::improved} provides a scheme for $n=2^T-1$. Section~\ref{sec::BipartiteGraphRep} discusses special instances of the problem for a general $n$, while Section~\ref{sec::algorithms} presents upper bounds and algorithms. Section~\ref{sec::evaluation} provides numerical evaluations, and finally Section~\ref{sec::related} discusses related work.

\section{Problem Formulation and Previous Results}
\label{sec::setup}

\smallskip
\noindent\textbf{Notation.} 
Calligraphic letters indicate sets;
boldface lower case letters denote vectors and boldface upper case letters indicate matrices;
$|\Set{X}|$ is the cardinality of $\Set{X}$;
$[n]$ is the set of integers $\{1,\cdots,n\}$;
$\emptyset$ is the empty set;
for all $x \in \mathbb{R}$, the floor and ceiling functions are denoted with $\lfloor x \rfloor$ and $\lceil x \rceil$, respectively;
$\mathbf{0}_{j}$ is the all-zero row vector of dimension $j$;
$\mathbf{0}_{i \times j}$ is the all-zero matrix of dimension $i \times j$;
$\mathbf{1}_{j}$ denotes a row vector of dimension $j$ of all ones;
logarithms are in base 2.

\smallskip
{\noindent\textbf{Index Coding.} We consider a setup similar to the one in \cite{karmoose2017preserving}. We assume an index coding instance, where a server has a database $\Set{B}=\left \{ \mathbf{b}_{\Set{M}} \right \}$ of $m$ messages, with $\Set{M} = [m]$ being the set of message indices, and all messages $\mathbf{b}_j \in \mathbb{F}_{2}^{F}, j \in \Set{M},$ are $F$-long strings.
The server is connected through a broadcast channel to a set of clients $\Set{C}=\left \{ c_{\Set{N}} \right \}$, where $\Set{N} = [n]$ is the set of client indices, and $m \geq n$. Each client {$c_i, i \in \Set{N},$} has a subset of the messages {$\left \{ \mathbf{b}_{\Set{S}_i}\right \}$, with $\Set{S}_i \subset \Set{M}$,} as side information and requests a new message {$\mathbf{b}_{q_i}$} with $q_i \in \Set{M} \setminus \Set{S}_i$ that {it} does not have. 
A \textit{linear index code} solution to the index coding instance is a designed set of broadcast transmissions that are linear combinations of the messages in {$\Set{B}$}. The linear index code can be represented as $\mathbf{A} \mathbf{B} = \mathbf{Y}$, where $\mathbf{A} \in \mathbb{F}_2^{T \times m}$ is the coding matrix, $\mathbf{B} \in \mathbb{F}_2^{m \times F}$ is the matrix of all the messages and $\mathbf{Y} \in \mathbb{F}_2^{T \times F}$ is the resulting matrix of linear combinations. Upon receiving these transmissions, client $c_i{, i \in \Set{N},}$ employs linear decoding to retrieve $\mathbf{b}_{q_i}$. A linear index code with the minimum number of transmissions is called an \textit{optimal} linear index code.}

\smallskip
\noindent\textbf{Problem Formulation.}
{Designing the optimal linear index code is an NP-Hard problem, and therefore various algorithms exist for designing sub-optimal linear index codes (see Section~\ref{sec::related}). In this work, we are concerned with designing linear index codes that maintain higher privacy levels for the requests of clients. Our approach is based on using $k$-limited-access schemes~\cite{karmoose2017preserving}: given a coding matrix $\mathbf{A}$ of rank $T$, we wish to create an alternative index code $\mathbf{A}_k = \mathbf{P} \mathbf{A}$, where $\mathbf{P} \in \mathbb{F}_2^{T_k \times T}$ is to be designed such that client $c_i, i \in \Set{N}$ can retrieve $b_{q_i}$ using \textit{at most} $k$ vectors of $\mathbf{A}_k$, where $1 \leq k \leq T$. The value of $T_k$ represents the number of transmissions associated with the alternative index code $\mathbf{A}_k$, and therefore our goal is to design $\mathbf{P}$ with minimum $T_k$. In order to create such a linear index code, we note that the coding matrix $\mathbf{A}$ allows client $c_i, i \in \Set{N}$ to retrieve $b_{q_i}$ by a linear decoding operation expressed as $\mathbf{d}_i \mathbf{A} \mathbf{B} = \mathbf{d}_i \mathbf{Y}$, where $\mathbf{d}_i \in \mathbb{F}_2^T$ is the decoding row vector of $c_i$. The resulting vector $\mathbf{g}_i = \mathbf{d}_i \mathbf{A}$ possesses certain properties which allows $c_i$ to decode $b_{q_i}$ using $b_{\Set{S}_i}$~\cite{bar2011index}. 
Therefore, an alternative index code $\mathbf{A}_k$ would still allow client $c_i$ to decode $b_{q_i}$ if it is able to reconstruct $\mathbf{g}_i$ using $\mathbf{A}_k$.
Our problem can therefore be stated as follows:
{\it Given $\mathbf{g}_i, i \in \Set{N}$, can we design a matrix $\mathbf{P}$, with $T_k$ as small as possible, such that $\mathbf{g}_i, i \in \Set{N}$ can be reconstructed by adding at most $k$ vectors out of $\mathbf{A}_k$?} Note that, by definition, $\mathbf{g}_i, i \in \Set{N}$ lie in the row span of $\mathbf{A}$. Since the rank of $\mathbf{A}$ is $T$, the maximum number of distinct $\mathbf{g}_i$ vectors is $2^T-1$. Therefore, without loss of generality, we assume that $n \leq 2^T-1$. We refer to the case where $n = 2^T-1$ as {\it full-space covering}, and to the case where $n < 2^T-1$ as {\it partial-space covering}.

Our previous work in~\cite{karmoose2017preserving} provided a lower bound on the minimum value of $T_k$, which we restate here for convenience.

\begin{lemma}
 \label{lem::lowerbound}\cite[Theorem III.1]{karmoose2017preserving}
 Given an index coding matrix $\mathbf{A} \in \mathbb{F}_2^{T \times m}$ with $T \geq 2$, it is possible to transform it into ${\mathbf{A}_{k}} = \mathbf{P} \mathbf{A}$ with $\mathbf{P} \in \mathbb{F}_2^{T_k \times T}$, such that each client can recover its request by combining at most $k$ rows of it, if and only if
\begin{align}
 T_k \geq T^\star \!=\! \min \left\{T_k : \sum\limits_{i=1}^{k} {T_k \choose i} \!\geq\! n \right\} \stackrel{(a)}{\geq} T^{\text{LB}} \!=\! \frac{2^{\frac{T\!-\!1}{k}} k^{\frac{k-1}{k}}}{e}, \label{eq::lb}
\end{align}
where $(a)$ holds when $n = 2^T-1$ and $k < \lceil T/2 \rceil$.
\end{lemma}

In addition, \cite[Theorem III.1]{karmoose2017preserving} provided a construction of a matrix $\mathbf{P}$ for which $T_k$ is shown to have an exponent that is order-optimal for the full-space covering case and for some regimes of $k$. Differently, one contribution in this paper is a matrix construction that is order-optimal for any value of $1 \leq k < \left\lceil T/2 \right\rceil$\footnote{The case $\left\lceil T/2 \right\rceil \leq k < T$ was solved in~\cite{karmoose2017preserving}, where we showed that $T_k= \min \{ T+1,n\}$.}. This is described in the next section.
}

\section{Improved Scheme for Full-Space Covering}
\label{sec::improved}
Here we provide a novel scheme for the full-space covering case (i.e., $n = 2^T-1$). This new scheme is order-optimal in the number of transmissions for the case when $ 1 \leq k < \left\lceil \frac{T}{2} \right\rceil$. This provides an improvement over the scheme presented in \cite[Theorem III.1]{karmoose2017preserving}.

\begin{theorem}
\label{thm::fullspacescheme}
 For $n = 2^T-1$ and $ 1 \leq k < \left\lceil \frac{T}{2} \right\rceil$ we have
 \begin{equation}
 \label{eq::fullspacescheme}
  T_k \leq 2^{\left\lceil \frac{T}{k} \right\rceil}k.
 \end{equation}
\end{theorem}
Before providing the proof for Theorem \ref{thm::fullspacescheme}, which shows how the scheme is constructed, we analyze the performance of the scheme in comparison to the lower bound in \eqref{eq::lb}. We do so in the next lemma (proof is in Appendix 1).
\begin{lemma}
 \label{lem::orderoptimal}
 For $ 1 \leq k < \left\lceil \frac{T}{2} \right\rceil$, we have $T_k = \Theta(2^{\frac{T}{k}}k)$.
\end{lemma}   

 \begin{figure}
  \centering
  \includegraphics[width=3in]{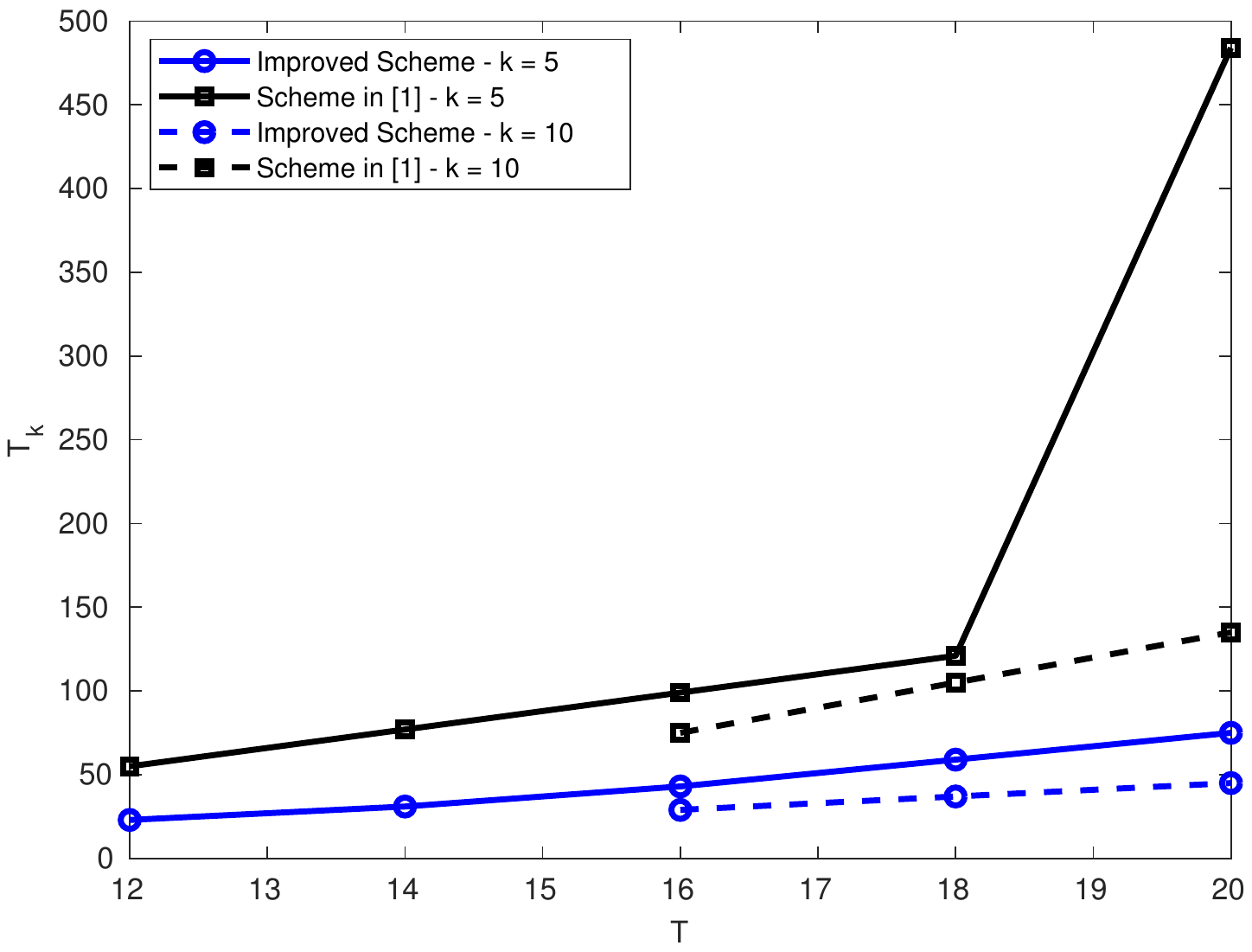}
  \caption{Comparison between the scheme in Theorem \ref{thm::fullspacescheme} and in~\cite{karmoose2017preserving}}
  \label{fig::compfullspace}
 \end{figure}
 
  The main difference between Scheme-1 (in Theorem~\ref{thm::fullspacescheme}) and Scheme-2 (in~\cite{karmoose2017preserving}) is as follows. 
Both schemes are designed {by: (i)} breaking the {binary} vector {of length $T$} into parts, {(ii)} providing all possible {non-zero binary} vectors that correspond to each part, and {(iii) combining} the solutions to {reconstruct the original vector. However,} 
the two schemes {differ} in the following: 1) Scheme-1 splits the vector into {\it larger} but {\it fewer} parts {than Scheme-2,} and 2) Scheme-1 aggregates the {solutions} {\it additively} while {Scheme-2} aggregates {them} {\it multiplicatively}. While it is indeed true that providing all possible vectors for the parts in Scheme-1 would lead to larger partial solutions than those in Scheme-2, aggregating those solutions additively eventually leads to a smaller number of vectors than in Scheme-2. Figure~\ref{fig::compfullspace} shows a comparison between the improved scheme proposed in~\ref{thm::fullspacescheme} and its counterpart in~\cite{karmoose2017preserving} for full-space covering.

 The remainder of this section proves Theorem \ref{thm::fullspacescheme} by showing how the scheme works (i.e. how $\mathbf{A}_k$ is constructed).

\medskip
 
\noindent
 \textbf{Example:} We first show how the scheme is constructed via a small example, where $T = 8$ and $k = 3$. The idea is that, to reconstruct a vector $\mathbf{v} \in \mathbb{F}_2^8$, we treat it as $k=3$ disjoint parts; the first $2$ are of length $\left\lceil \frac{T}{k} \right\rceil = 3$ and the remaining part is of length $T - (k-1)\left\lceil \frac{T}{k} \right\rceil = 2$. We then construct $\mathbf{A}_k$ as $k = 3$ disjoint sections, where each section allows us to reconstruct one part of the vector. Specifically, we construct $\mathbf{A}_k$ as 
 
 \begin{equation}
 \nonumber
  \mathbf{A}_k = \left[ \begin{matrix}
                       \bar{\mathbf{B}}_1 & \mathbf{0}_{7 \times 3} & \mathbf{0}_{7 \times 2} \\
                       \mathbf{0}_{7 \times 3} & \bar{\mathbf{B}}_2 & \mathbf{0}_{7 \times 2} \\
                       \mathbf{0}_{3 \times 3} &  \mathbf{0}_{3 \times 3} & \bar{\mathbf{B}}_3
                      \end{matrix}
 \right],
 \end{equation}
where 
%
\begin{equation}
 \nonumber
 \bar{\mathbf{B}}_1 = \bar{\mathbf{B}}_2 = \left[ \begin{matrix}
                                                   0 & 0 & 1 \\ 0 & 1 & 0 \\ 0 & 1 & 1 \\ & \vdots & \\ 1 & 1 & 1
                                                  \end{matrix}
 \right], \quad \bar{\mathbf{B}}_3 = \left[ \begin{matrix}
                                             0 & 1 \\ 1 & 0 \\ 1 & 1
                                            \end{matrix}
 \right].
\end{equation}
Then any vector $\mathbf{v}$ can be reconstructed by picking at most $k$ vectors out of $\mathbf{A}_k$, one from each section. For example, let $\mathbf{v} = \left[ 0 \: 1 \: 0 \: 0 \: 1 \: 1 \: 1 \: 0 \right]$. Then this vector can be reconstructed by adding vectors number $2$, $10$ and $16$ from $\mathbf{A}_k$.

 \medskip
 
\noindent
 \textbf{Proof of Theorem \ref{thm::fullspacescheme}:}
 Let $T_{\text{rem}} = T - (k-1)\left\lceil \frac{T}{k}\right\rceil $. Then we can write 
%
\begin{equation}
\nonumber
 \mathbf{A}_k = \left[ \begin{matrix}
 \mathbf{B}_1 \\
 \mathbf{B}_2 \\
 \vdots \\
 \mathbf{B}_{k}\end{matrix}  \right],
\end{equation}

where, for $i \in [k-1]$, the matrix $\mathbf{B}_i$, of dimension $b_i \times T$, is constructed as follows
\begin{equation}
 \nonumber
 \mathbf{B}_i = \left[ \begin{matrix} \mathbf{0}_{b_i \times (i-1) \left\lceil \frac{T}{k} \right\rceil} & \bar{\mathbf{B}}_i &  \mathbf{0}_{b_i \times (k-1-i) \left\lceil \frac{T}{k} \right\rceil} & \mathbf{0}_{b_i \times T_{\text{rem}}} \end{matrix} \right],
\end{equation}
where $\bar{\mathbf{B}}_i$, of dimension $b_i \times \left\lceil \frac{T}{k} \right\rceil$, has as rows all non-zero vectors of dimension $\left\lceil \frac{T}{k} \right\rceil$. Therefore we have $b_i = 2^{\left\lceil \frac{T}{k} \right\rceil }-1$.

Similarly, the matrix $\mathbf{B}_k$, of dimension $b_k \times T$, is constructed as follows
\begin{equation}
 \nonumber
 \mathbf{B}_k = \left[ \begin{matrix} \mathbf{0}_{b_k \times (k-1) \left\lceil \frac{T}{k} \right\rceil} & \bar{\mathbf{B}}_k \end{matrix} \right],
\end{equation}
where $\bar{\mathbf{B}}_k$, of dimension $b_k \times T_{\text{rem}}$, has as rows all non-zero vectors of dimension $T_{\text{rem}}$. Therefore we have $b_k = 2^{T_{\text{rem}}}-1$.

In other words, the matrix $\mathbf{A}_k$ is constructed as a block-diagonal matrix, with the diagonal elements being $\bar{\mathbf{B}}_i$ for all $i \in [k]$. Therefore equation \eqref{eq::fullspacescheme} holds by computing 
\begin{equation}
\nonumber
 T_k = \sum\limits_{i=1}^{k} b_i = (k-1)\left(2^{\left\lceil \frac{T}{k} \right\rceil} -1 \right) + 2^{T_{\text{rem}}} - 1 \leq k2^{\left\lceil \frac{T}{k} \right\rceil},
\end{equation}
which follows by noting that $T_{\text{rem}} \leq \left\lceil \frac{T}{k} \right\rceil$.

What remains is to show that any vector $\mathbf{v} \in \mathbb{F}_2^T$ can be reconstructed by adding at most $k$ vectors of $\mathbf{A}_k$. To show this, we can express it as $\mathbf{v} = \left[\mathbf{v}_1 \: \cdots \: \mathbf{v}_k \right]$ where $\mathbf{v}_i, i \in [k-1]$ are parts of the vector $\mathbf{v}$ each of length $ \left\lceil \frac{T}{k} \right\rceil $, while $\mathbf{v}_k$ is the last part of $\mathbf{v}$ of length $T_{\text{rem}}$. Then we can write 
\begin{equation}
 \nonumber
 \mathbf{v} = \sum\limits_{i=1}^k \bar{\mathbf{v}}_i = \sum\limits_{i \in \mathcal{K}(\mathbf{v})} \bar{\mathbf{v}}_i,
\end{equation} 
where $\bar{\mathbf{v}}_i = \left[ \mathbf{0}_{(i-1) \left\lceil \frac{T}{k} \right\rceil} \quad \mathbf{v}_i \quad \mathbf{0}_{(k-1-i) \left\lceil \frac{T}{k} \right\rceil} \: \mathbf{0}_{T_{\text{rem}}} \right]$ for $i \in [k-1]$, $\bar{\mathbf{v}}_k = \left[ \mathbf{0}_{(k-1) \left\lceil \frac{T}{k} \right\rceil} \quad \mathbf{v}_k \right]$ and $\mathcal{K}(\mathbf{v}) \subseteq [k]$ is the set of indices for which $\mathbf{v}_i$ is not all-zero. Then, according to the construction of $\mathbf{A}_k$, for all $i \in \mathcal{K}(\mathbf{v})$, the corresponding vector $\mathbf{v}_i$ is one of the rows in $\mathbf{B}_i$, and therefore we can construct $\mathbf{v}$ by at most $k$ vectors of $\mathbf{A}_k$. \hfill$\blacksquare$

\section{Partial-Space Covering}
\label{sec::BipartiteGraphRep}
\begin{figure}
  \centering
  \begin{minipage}{0.49\columnwidth}
  \centering
  \includegraphics[width=0.95\textwidth]{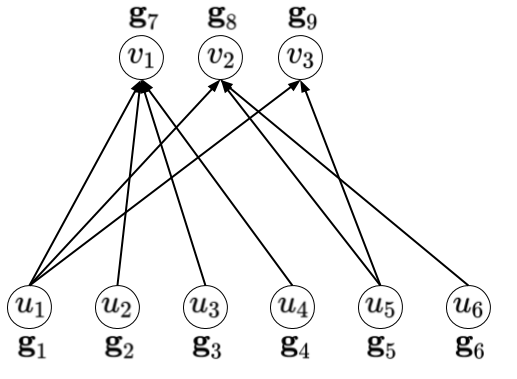}
  \caption{Bipartite graph representation.}
  \label{fig::bipartitegraphrep}
    \end{minipage}
      \begin{minipage}{0.49\columnwidth}
  \centering
  \includegraphics[width=0.95\textwidth]{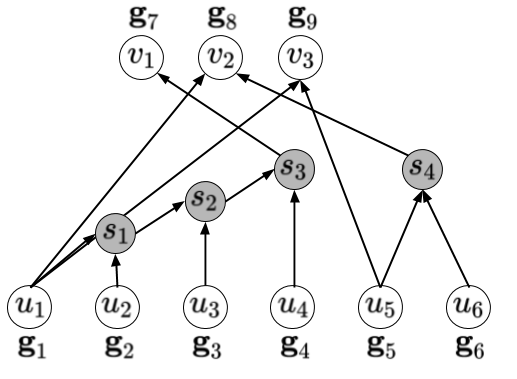}
  \caption{Optimal representation when $k=2$.}
  \label{fig::bipartitegraphrepsol}
    \end{minipage}
\end{figure}

Here we study some specific instances of the problem, which we will later use in our algorithms. We first represent the problem through a bipartite graph as follows. We assume that the rank of the matrix $\mathbf{G}$ is $T$. Then, there exists a set of $T$ linearly independent vectors in $\mathbf{G}$; without loss of generality, denote them as $\mathbf{g}_1$ to $\mathbf{g}_T$.
We can then represent the problem as a bipartite graph $(\Set{U} \cup \Set{V}, \Set{E})$ with $| \Set{U}| = T$ and $| \Set{V}| = n-T$, where $u_i \in \Set{U}$ represents vector $\mathbf{g}_{i}$ for $i \in [T]$, $v_i \in \Set{V}$ represents vector $\mathbf{g}_{i+T}$ for $i \in [n-T]$, and an edge exists from node $u_i$ to node $v_j$ if $\mathbf{g}_i$ is one of the component vectors of $\mathbf{g}_{j+T}$. Figure~\ref{fig::bipartitegraphrep} shows an example of such graph, where $n = 9$ and $T = 6$. 
For instance, $v_1$ (i.e., $\mathbf{g}_7$) can be reconstructed by adding $u_i, i \in [4]$.
Given a node $s$ in the graph, we refer to the sets $\Set{O}_s$ and $\Set{I}_s$ as the {\it outbound} and {\it inbound} sets of $s$ respectively: the inbound set contains the nodes which have edges outgoing to node $s$, and the outbound set contains the nodes to which node $s$ has outgoing edges. 
For instance, with reference to Figure~\ref{fig::bipartitegraphrep}, $\Set{O}_{u_1} = \{v_1,v_2,v_3 \}$ and $\Set{I}_{v_1} = \{u_1,u_2,u_3,u_4 \}$.
For this particular example, there exists a scheme with $T_2 = 6$ which can reconstruct any vector with at most $k=2$ additions. The matrix $\mathbf{A}_2$ which corresponds to this solution consists of the following vectors:
\begin{equation}
 \mathbf{A}_2 = \left[ \begin{matrix}
                        \mathbf{g}_1 \\
                        \mathbf{g}_1 + \mathbf{g}_2 \\
                        \mathbf{g}_1 + \mathbf{g}_2 + \mathbf{g}_3 \\
                        \mathbf{g}_1 + \mathbf{g}_2 + \mathbf{g}_3 + \mathbf{g}_4\\
                        \mathbf{g}_5 \\
                        \mathbf{g}_5 + \mathbf{g}_6
                       \end{matrix}
\right].
\end{equation}
It is not hard to see that each vector in $\mathbf{G}$ can be reconstructed by adding at most $2$ vectors in $\mathbf{A}_2$. The vectors in $\mathbf{A}_2$ that are not in $\mathbf{G}$ can be aptly represented as intermediate nodes on the previously described bipartite graph, which are shown in Figure \ref{fig::bipartitegraphrepsol} as highlighted nodes. Each added node represents a new vector, which is the sum of the vectors for the nodes in its inbound set. We refer to the process of adding these intermediate nodes as creating a {\it branch}, which is defined next.

\begin{definition} Given an ordered set $\Set{S} = \{s_1, \: \cdots, \: s_S \}$ of nodes,
where $s_i$ preceeds $s_{i+1}$ for $i \in [S-1]$, 
a {\it branch on $\Set{S}$} is a set $\Set{S}^\prime = \{ s^\prime_1, \: \cdots, \: s^\prime_{S-1} \}$ of $S-1$ intermediate nodes added to the graph with the following connections: node $s^\prime_1$ has two incoming edges from $s_1$ and $s_2$, and for $i \in [S-1] \setminus \{1\}$, $s^\prime_i$ has two incoming edges from nodes $s^\prime_{i-1}$ and $s_{i+1}$.
\end{definition}

For the example in Figure \ref{fig::bipartitegraphrepsol}, we created branches on two ordered sets, $\Set{S}_1 = \{u_1, \: u_2, \: u_3, \: u_4 \}$ and $\Set{S}_2 = \{u_5, \: u_6 \}$. Once the branch is added, we can change the {connections} of the nodes in $\Set{V}$ in accordance to the added vectors. For the example in Figure \ref{fig::bipartitegraphrepsol}, we can replace $u_{[4]}$ in $\Set{I}_{v_1}$ with only $s_{3}$.

Using this representation, we have the following lemma.

\begin{lemma}
\label{lem::subsetlem}
 If $\mathcal{O}_{u_{i_T}} \subseteq \mathcal{O}_{u_{i_{T-1}}} \subseteq \cdots \subseteq \mathcal{O}_{u_{i_1}}$ for some permutation $i_1, \cdots, i_T$ of $[T]$, then this instance can be solved by exactly $T$ transmissions for any $k \geq 2$.
\end{lemma}

\begin{Pf}
 One solution of such instance would involve creating a branch on the set $\Set{S} = \{u_{i_T}, \: u_{i_{T-1}}, \: \cdots, \: u_{i_1} \}$.
 The scheme used would have the matrix $\mathbf{A}_2$ with its $t$-th row $\mathbf{a}_t = \sum\limits_{\ell = 1}^{t} \mathbf{g}_{i_\ell}$ for $t \in [T]$. Note that $\mathbf{g}_{i_1} = \mathbf{a}_{1}$ and $\mathbf{a}_{t} + \mathbf{a}_{t-1} = \mathbf{g}_{i_t}$ for all $t \in [T]\setminus\{1\}$. 
 Moreover, for $j \in [n]\setminus [T]$, if $v_{j-T} \in \mathcal{O}_{u_{i_t}}$ for some $i_t$, then $ v_{j-T} \in \mathcal{O}_{u_{i_\ell}}$ for all $\ell \leq t$. 
If we let $t$ be the maximum index for which $v_{j-T} \in \mathcal{O}_{u_{i_t}}$, then we have $\Set{I}_{v_{j-T}} = \{u_{i_1}, \: \cdots, \: u_{i_t}\}$, and so we get $\mathbf{g}_{j} = \sum\limits_{\ell = 1}^t \mathbf{g}_{i_\ell} = \mathbf{a}_t$.
\end{Pf}

\begin{corollary}
\label{cor::circuit}
 For $\mathbf{G} \in \mathbb{F}_2^{n \times T}$ of rank $T$, if $n = T+1$, then this instance can be solved in $T$ transmissions for any $k \geq 2$.
\end{corollary}
\begin{Pf}
Without loss of generality, let $\mathbf{g}_{[T]}$ be a set of linearly independent vectors of $\mathbf{G}$. 
Then we have $\Set{O}_{u_i} = \{v_1\}$ for $i \in \Set{I}_{v_1}$ and $\Set{O}_{u_j} = \emptyset$ for $j \in [T]\setminus \Set{I}_{v_1}$. Thus, from Lemma~\ref{lem::subsetlem}, this instance can be solved in $T$ transmissions.
\end{Pf}

\section{Algorithms for General Instances}
\label{sec::algorithms}

\subsection{Successive Circuit Removing (SCR) algorithm} 
Our first proposed algorithm is based on Corollary~\ref{cor::circuit}, which can be interpreted as follows: any matrix $\mathbf{G}$ of $r+1$ vectors and rank $r$ can be reconstructed by a corresponding $\mathbf{A}_2$ matrix with $r$ rows. 
We denote this collection of vectors as a \textit{circuit}\footnote{This is in accordance to the definition of a circuit for a matroid\cite{oxley2006matroid}.}. Our algorithm works for the case $k = 2^q$, for some integer $q$. We first describe SCR for the case where $q=1$, and then extend it to a general $q$.
For $q=1$, the algorithm works as follows:

\medskip

\noindent $1)$ \textit{Circuit Finding:} find a set of vectors of $\mathbf{G}$ that form a circuit of small size. Denote the size of this circuit as $r+1$.\\
\noindent $2)$ \textit{Matrix Update:} apply Corollary \ref{cor::circuit} to find a set of $r$ vectors that can optimally reconstruct the circuit by adding at most $k=2$ of them, and add this set to $\mathbf{A}_2$.\\
\noindent $3)$ \textit{Circuit Removing:} update $\mathbf{G}$ by removing the circuit. Repeat the first two steps until the matrix $\mathbf{G}$ is of size $T^\prime \times T$ and of rank $T^\prime$, where $T^\prime \leq T$. Then add these vectors to $\mathbf{A}_2$.

\medskip

Once SCR is executed, the output is a matrix $\mathbf{A}_2$ such that any vector in $\mathbf{G}$ can be reconstructed by adding at most $k=2$ vectors of $\mathbf{A}_2$. Consider now the case where $q=2$ (i.e., $k=4$) for example. In this case, a second application of SCR on the matrix $\mathbf{A}_2$ would yield another matrix, denoted as $\mathbf{A}_4$, such that any row in $\mathbf{A}_2$ can be reconstructed by adding at most $2$ vectors of $\mathbf{A}_4$. Therefore any vector in $\mathbf{G}$ can now be reconstructed by adding at most $4$ vectors of $\mathbf{A}_4$. We can therefore extrapolate this idea for a general $q$ by successively applying SCR $q$ times on $\mathbf{G}$ to obtain $\mathbf{A}_k$, with $k=2^q$.

The following theorem gives a closed form characterization of the best and worst case performance of SCR.

\begin{theorem}
 \label{thm::SCR}
 Let $T_2^{\text{SRC}}$ be the number of vectors in $\mathbf{A}_k$ obtained via SCR. Then, for $k=2^q$ and integer $q$, we have
 \begin{equation}
  \label{eq::SRCPerf}
  \underbrace{f^{\text{Best}}(f^{\text{Best}}( \cdots f^{\text{Best}}(n)))}_{q \text{ times}} \!\leq\! T_q^{\text{SRC}} \!\leq \! \underbrace{f^{\text{Worst}}(f^{\text{Worst}}( \cdots f^{\text{Worst}}(n)))}_{q \text{ times}},
 \end{equation}
 where $f^{\text{Best}}(n) = 2\left\lfloor \frac{n}{3}\right\rfloor$ and $f^{\text{Worst}}(n) = T \left( \left\lfloor \frac{n}{T+1} \right\rfloor + 1\right)$.
\end{theorem}
\begin{Pf}
First we focus on the case $q = 1$.
The lower bound in~\eqref{eq::SRCPerf} corresponds to the best case when the matrix $\mathbf{G}$ can be partitioned into disjoint circuits of size $3$. In this case, if SRC finds one such circuit in each iteration, then each circuit is replaced with $2$ vectors in $\mathbf{A}_2$ according to Corollary~\ref{cor::circuit}. To obtain the upper bound, note that any collection of $T+1$ has at most $T$ independent vectors, and therefore contains a circuit of at most size $T+1$. Therefore, the upper bound corresponds to the case where the matrix $\mathbf{G}$ can be partitioned into circuits of size $T+1$ and an extra $T$ linearly independent vectors. In that case, the algorithm can go through each of these circuits, adding $T$ vectors to $\mathbf{A}_2$ for each of these circuits, and then add the last $T$ vectors in the last step of the algorithm. Finally, the bounds in \eqref{eq::SRCPerf} for a general $q$ can be proven by a successive repetition of the above arguments.
\end{Pf}

\subsection{Branch-Search heuristic}
A naive approach to determining the optimal matrix $\mathbf{A}_k$ is to consider the whole space $\mathbb{F}_2^T$, loop over all possible subsets of vectors of $\mathbb{F}_2^T$ and, for every subset, check if it can be used as a matrix $\mathbf{A}_k$. The minimum-size subset which can be used as $\mathbf{A}_k$ is indeed the optimal matrix. However, such algorithm requires in the worst case) $O\left (2^{2^T} \right )$ number of operations, which makes it prohibitively slow even for very small values of $T$. 
Instead, {the heuristic that we here propose} finds a matrix $\mathbf{A}_k$ more efficiently than the naive search scheme. The main idea behind the heuristic is based on providing a subset $\Set{R} \subset \mathbb{F}_2^T$ which is much smaller than $2^T$ and is guaranteed to have at least one solution. The heuristic then searches for a matrix $\mathbf{A}_k$ by looping over all possible subsets of $\Set{R}$. Our heuristic therefore consists of two sub-algorithms, namely Branch and Search. 
Branch takes as input $\mathbf{G}$, and produces as output a set of vectors $\Set{R}$ which contains at least one solution $\mathbf{A}_k$. The algorithm works as follows:

\medskip

\noindent 1) Find a set of $T$ vectors of $\mathbf{G}$ that are linearly independent. Denote this set as $\Set{B}$.\\
\noindent 2) Create a bipartite graph representation of $\mathbf{G}$ as discussed in Section~\ref{sec::BipartiteGraphRep}, using $\Set{B}$ as the independent vectors for $\Set{U}$.\\
\noindent 3) Pick the dependent node $v_i$ with the highest degree, and split ties arbitrarily. Denote by $\text{deg}(v_i)$ the degree of node $v_i$.\\
\noindent 4) Consider the inbound set $\Set{I}_{v_i}$, and sort its elements in a descending order according to their degrees. Without loss of generality, assume that this set of ordered independent nodes is $\Set{I}_{v_i} = \{u_1, \: u_2, \: \cdots , \: u_{\text{deg}(v_i)} \}$.\\
\noindent 5) Create a branch on $\Set{I}_{v_i}$. Denote the new branch nodes as $\{u^\star_1, \: u^\star_2, \: \cdots , \: u^\star_{\text{deg}(v_i)} \}$.\\
\noindent 6) Update the connections of all dependent nodes in accordance with the constructed branch. This is done as follows: for each node $v_j \in \Set{V}$ with $\text{deg}(v_j) \geq k$, if $\Set{I}_{v_j} \cap \Set{I}_{v_i}$ is of the form $\{u_1, \: u_2, \: \cdots, \: u_{\ell}\}$ for some $\ell \leq \text{deg}(v_i)$, then replace $\{u_1, \: u_2, \: \cdots, \: u_{\ell}\}$ in $\Set{I}_{v_j}$ with the single node $u^\star_{\ell}$.\\
\noindent 7) Repeat 3) to 6) until all nodes in $\Set{V}$ have degree at most $k$.

\medskip

The output $\Set{R}$ is the set of vectors corresponding to all nodes in the graph. The next theorem shows that $\Set{R}$ in fact contains one possible $\mathbf{A}_k$, and characterizes the performance of Branch.

\begin{theorem}
\label{thm::branch} (Proof in Appendix 2) 
 For a matrix $\mathbf{G}$ of dimension $n \times T$, (a) Branch produces a set $\Set{R}$ which contains at least one possible $\mathbf{A}_k$, (b) the worst-case time complexity $t_{\text{Branch}}$ of Branch is $O(n^2)$, and (c) $|\Set{R}| \leq (n-T)T$.
\end{theorem}

Let $t_{\text{Search}}$ be the worst-time complexity of the Search step in Branch-Search. Then the worst-case time complexity of Branch-Search is equal to $t_{\text{BS}} = t_{\text{Branch}} + t_{\text{Search}} \leq O(n^2) + 2^{|\Set{R}|} = O(n^2) + O(2^{nT}) = O(2^{nT})$, which is exponentially better than the complexity of naive search. Although our heuristic is still of exponential runtime complexity, we observe from numerical simulations that $|\Set{R}|$ is usually much less than $(n-T)T$. Moreover, we believe that there exist more efficient ways of searching through the set $\Set{R}$ to find a better solution $\mathbf{A}_k$, which is part of our ongoing investigation.

\section{Numerical Evaluation}
\label{sec::evaluation}

\begin{figure}
 \centering
   \includegraphics[width=0.933\columnwidth]{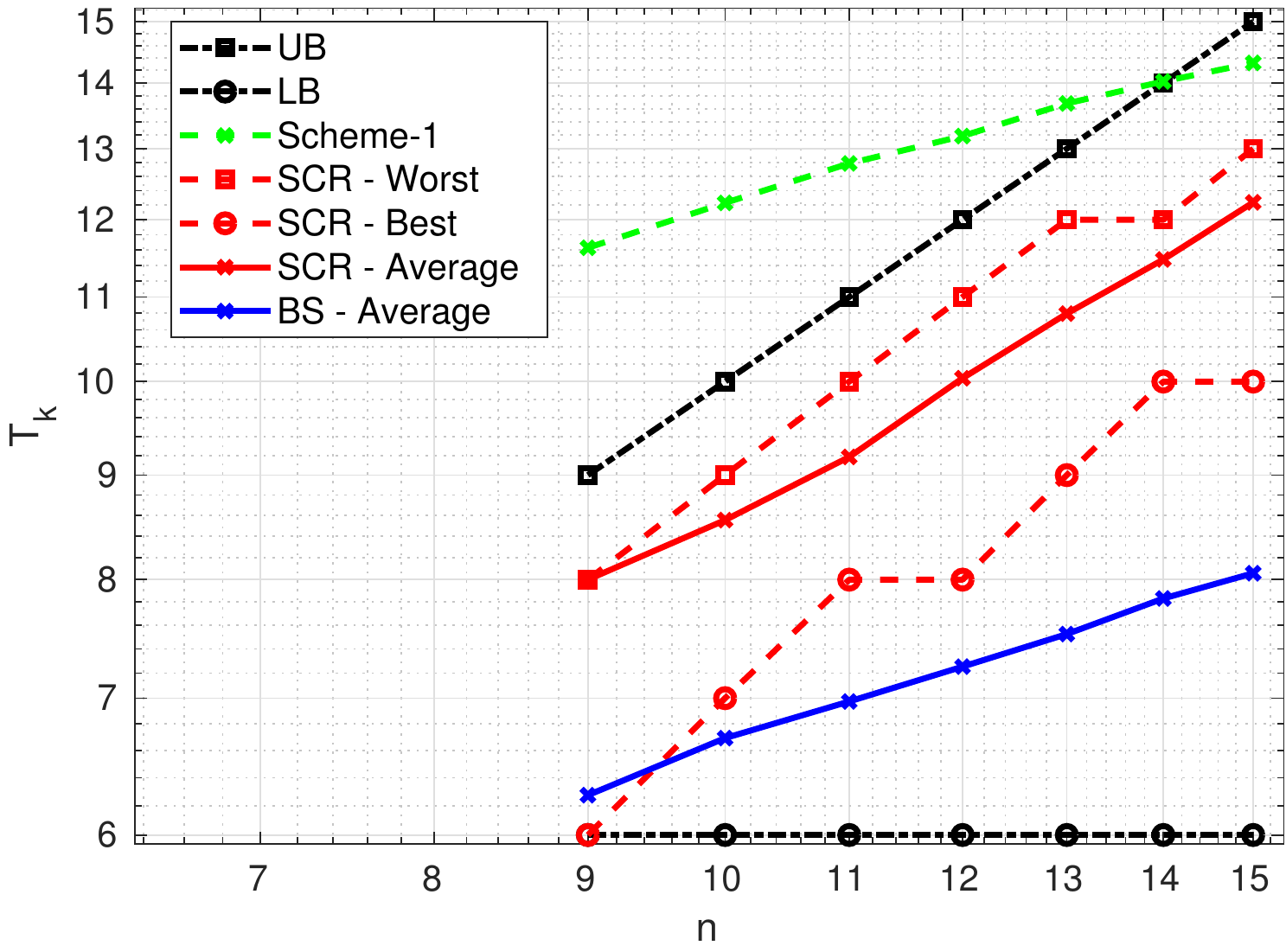}
 \caption{Numerical evaluation - $T = 6$, $k=2$.}
 \label{fig::algperf}
 \end{figure}
%

Here we evaluate the performance of our proposed schemes through numerical evaluations. Specifically, we assess the performance in terms of $T_k$ of the scheme in Theorem \ref{thm::fullspacescheme} (which we here refer to as Scheme-1), SCR and Branch-Search (labeled BS). {\karmoose We compare their performance against the lower bound in Lemma \ref{lem::lowerbound} (denoted by LB), and the upper bound of sending uncoded transmissions (denoted by UB).} For the case of partial-space covering, we adapt Scheme-1 in the following way: we first sort the columns of $\mathbf{G}$ in a decreasing order according to their weights (i.e., number of non-zero elements), then for the $i$-th section of length $\left\lceil T/k \right\rceil$, we fill $\mathbf{B}_i$, {\karmoose not with all non-zero vectors of length $\left\lceil T/k \right\rceil$ (as described in the proof of Theorem \ref{thm::fullspacescheme}), but only with all the vectors that appear for that section across all the $n$ vectors of $\mathbf{G}$}. This modification removes vectors from the matrix $\mathbf{A}_k$ that are not used by any vector in $\mathbf{G}$. For SCR, we evaluate its average performance as well as its upper and lower bound performance established in Theorem~\ref{thm::SCR}. For Branch-Search, we evaluate its average performance. Figure~\ref{fig::algperf} shows the performance of all the aforementioned schemes for $T=6$ and $k=2$. As can be seen, Scheme-1 does not perform well for small values of $n$. 
SCR consistently performs better than uncoded transmissions.
In addition, although the current implementation of SCR greedy searches for a small circuit to remove, more sophisticated algorithms for small circuit finding could potentially improve its performance. However, the bounds in~\eqref{eq::SRCPerf} suggest that the performance of SCR is asymptotically $O(n)$. Branch-Search appears to perform better than other schemes in the average sense. Our current investigation includes understanding its asymptotic behavior in the worst-case.


\section{Related Work} \label{sec::related}
The problem of protecting privacy was initially proposed to enable the disclosure of databases for public access, while maintaining the anonymity of the users~\cite{aggarwal2008general}.
In {\it Private Information Retrieval} (PIR)~\cite{chor1998private,banawan2016capacity}, clients ensure that no information about their requests is revealed to a set of malicious databases when they retrieve information from them. 
Similarly, the problem of {\it Oblivious Transfer} (OT)~\cite{mishra2014oblivious} establishes, by means of cryptographic techniques, two-way private connections between the clients and the server.

We were here interested in addressing privacy concerns within the framework of index coding. This problem differs from secure index coding~\cite{dau2012security}: our goal is to protect the clients from an eavesdropper who wishes to learn the {\it identities}, rather than the {\it contents}, of the requested messages.
Our initial work in~\cite{karmoose2017private} addressed the possibility of designing coding matrices that provide privacy guarantees for clients.
The solutions based on $k$-limited-access schemes proposed in~\cite{karmoose2017preserving} can be interpreted as finding overcomplete bases that allow {\it sparse} representation of vectors, which is closely related to dictionary learning~\cite{rubinstein2010dictionaries}. However, finding lossless representation of vectors forbids us from using the efficient dictionary learning algorithms.
\section*{Appendix 1}
 To prove Lemma \ref{lem::orderoptimal}, we have to show that $T_k(T) = O(2^{\frac{T}{k}}k)$ and $T_k(T) = \Omega(2^{\frac{T}{k}}k)$. The notation $T_k(T)$ is to explicate that we are interested in the limiting behavior of $T_k$ as $T$ varies.  \\
 
 \noindent\underline{$T_k(T) = O(2^{\frac{T}{k}}k)$:} Let $c = 2$, then we have $2^{\frac{T}{k}+1}k = c \cdot 2^{\frac{T}{k}}k$. Therefore, for all $T \geq 1$, we can write
 \begin{equation}
  \nonumber
  T_k(T) \leq 2^{\left\lceil \frac{T}{k} \right\rceil}k \leq 2^{\frac{T}{k}+1}k = c \cdot 2^{\frac{T}{k}}k,
 \end{equation}
which proves the first part.\\

 \noindent\underline{$T_k(T) = \Omega(2^{\frac{T}{k}}k)$:} Proving $T_k(T) = \Omega(2^{\frac{T}{k}}k)$ is equivalent to proving that $2^{\frac{T}{k}}k = O(T_k(T))$ \cite[Definition 2.1]{arora2009computational}. 
 Let $c = e^{2/e}$. Then, for all $T \geq 1$, we have
  \begin{align}
  \nonumber
  2^{\frac{T}{k}} k &= (2k)^{1/k} \cdot 2^{\frac{T-1}{k}} k^{\frac{k-1}{k}} \\ 
		    &\stackrel{(a)}{\leq} e^{2/e} \cdot 2^{\frac{T-1}{k}} k^{\frac{k-1}{k}} \\
		    &= c \cdot 2^{\frac{T-1}{k}} k^{\frac{k-1}{k}} \\
		    &\stackrel{(b)}{=} c \cdot e \cdot T_{\text{LB}} \leq c \cdot e \cdot T_k(T)
 \end{align}
 where $(a)$ follows by noting that $f(k) = (2k)^{1/k} \leq e^{\frac{2}{e}}$ and $(b)$ follows by noting \eqref{eq::lb}. This proves the second part.
 \hfill$\blacksquare$
\section*{Appendix 2}
To see (a), note that the algorithm terminates when all dependent nodes have degrees $k$ or less. In every iteration of the algorithm, at least one dependent node is updated and its degree is reduced to $1$. Therefore the algorithm is guaranteed to terminate. Since all dependent nodes have degrees $k$ or less, then their corresponding vectors can be reconstructed by at most $k$ vectors in $\Set{R}$. Therefore, the set $\Set{R}$ contains one possible solution of $\mathbf{A}_k$.

To prove (b), the worst-case runtime of Branch corresponds to going over all nodes in $\Set{V}$, creating a branch for each one. For the $i$-th node considered by Branch, the algorithm would update the dependencies of all dependent nodes with degrees greater than $k$, which are at most $n-i$ nodes. Therefore $ t_{\text{Branch}} = \sum\limits_{i=0}^{n-1} (n-i) = n(n-1) = O(n^2)$.
 
 To prove (c), note that $|\Set{R}|$ is equal to the total number of nodes in all branches created by the algorithm. Therefore we can write $|\Set{R}| \leq \sum\limits_{v_i \in \Set{V}} \text{deg}(v_i) \leq (n-T)T = O(nT)$. \hfill $\blacksquare$

\bibliographystyle{IEEEtran}
\bibliography{Bib}

\end{document}